# Laser fragmentation of Aluminum nanoparticles in liquid isopropanol

Valery V. Smirnov[1], Margarita I. Zhilnikova[1,2], Ekaterina V. Barmina[1], Georgy A. Shafeev[1,3], Vitaly D. Kobtsev[4], Sergey A. Kostritsa[4], Svetlana M. Pridvorova[5]

1 - Prokhorov General Physics Institute of the Russian Academy of Sciences, 38, Vavilov street, 119991 Moscow Russian Federation

2 - Moscow Institute of Physics and Technology (National Research University), 9, Institutsky lane, Dolgoprudny, Moscow region, Russian Federation

3 - National Research Nuclear University MEPhI (Moscow Engineering Physics Institute), 31, Kashirskoye highway, 115409, Moscow, Russian Federation

4 - Central Institute of Aviation Motors, Aviamotornaya St. 2, Moscow 111116, Russian Federation

5 - A.N. Bach Institute of Biochemistry, Federal Research Center «Fundamentals of Biotechnology» of Russian Academy of Science, 33, Leninsky prospect, 119071, Moscow, Russian Federation

Key words: Hydrocarbon and composite fuels, laser fragmentation, Al nanoparticles, liquid

Metallic particles such as Aluminum possess rather high heat of combustion and enable to release large energy per unit volume. Properties of hydrocarbon fuels with addition of partially oxidized Al nanoparticles (NPs) was explored in [1,2]. The experimental research was carried out in [1] on the burning of the suspension of Al NPs of 50 nm in size in biofuel (ethanol). It was found that heat release under burning of ethanol scales almost linearly with the concentration of Al NPs. Volume fraction of Al NPs in the range of 1-3% does not affect the heat of combustion, while higher content of Al NPs up to 10% leads to the increase of volume heat release up to 15%.

In previous study Al NPs free of oxide layer were used as additive to n-decane $C_{10}H_{22}$. Such NPs were prepared in A.V. Luikov Heat and Mass Transfer Institute of National Academy of Science of Belarus by the method of plasma decomposition of triethylaluminum without presence of oxygen, followed by rapid cooling of the decomposition products by liquid hydrocarbon fuel [3]. It was experimentally found that at mass content of Al NPs in n-decane of 2.5% at some distance from the nozzle the increase of flame temperature was up to 450 K compared to pure n-decane. This was ascribed to the difference in combustion rates of the fuel components. At the same time, the expected difference in temperatures due to heat release of the composite Al NPs-decane fuel is of 100 – 150 K.

Several groups reported the increase of combustion rate and the decrease of ignition delay for some composite hydrocarbon fuels with Al NPs [3, 4, 5]. Therefore, the explorations of fuels with nano-additives attracted significant interest.

Alternative approach of synthesis of composite hydrocarbons with Al NPs consists in laser ablation of an Al target in liquid hydrocarbons [6-8]. Al NPs generated by this method are covered by layer of natural alumina. Further laser exposure of Al nanoparticles in liquid leads to the reduction of their size due to melting of individual NPs in the laser beam [9]. Molten NPs undergo hydrodynamic instability and are fragmented by high pressure vapors of the surrounding liquid. Laser ablation of Al target and laser fragmentation of Al NPs is conducted in liquid hydrocarbons due to its high reactivity. Al NPs can be prepared by laser fragmentation of Al micro- or nano-powder in water-free organic liquid. In this work we use nanoparticles of Al synthesized by electro-explosion of Al wires in vacuum (Alex) for further laser fragmentation in liquid isopropanol.

The size of Alex NPs exceeds 100 nm, and their suspension in isopropanol quickly sediments. To use them as an additive to hydrocarbon fuel their size should be reduced. Laser fragmentation was carried out using the radiation of a fiber ytterbium laser with pulse duration of 100 nanosecond at wavelength 1060-1070 nm and repetition rate of 20 kHz. Direct fragmentation of Al NPs in n-decane is accompanied by rapid decomposition of the hydrocarbon to elementary carbon. Carbon NPs strongly absorb laser radiation, and the liquid rapidly becomes opaque preventing thus further laser fragmentation of NPs. For this reason, the experiments on laser fragmentation of Alex were carried out using isopropanol as a liquid. This hydrocarbon also undergoes decomposition to elemental carbon with simultaneous emission of molecular $H_2$ [10]. However, the fraction of carbon nanoparticles in case of laser exposure of isopropanol is much lower than from n-decane, and the liquid remains transparent for laser radiation longer. Isopropanol is easily miscible with n-decane, and colloidal solution of Al NPs in it then can be added to n-decane in necessary proportion.

Fixed amount of Al NPs at concentration of order of 100 μg/ml was dispersed in isopropanol of total volume of 5-6 ml. The suspension of Al NPs was irradiated by a focused laser beam directed from bottom to top through a glass window of the cell transparent for laser radiation. The cell was continuously cooled with flowing water. The laser beam waist was positioned in the suspension above the window. The liquid was purged with $H_2$ supplied by hydrogen generator to substitute other gases dissolved in isopropanol [7]. Purging with $H_2$ also helps better agitate the suspended Al NPs. Laser beam was scanned across the window using a galvo-mirror system and F-theta objective. Estimated diameter of the laser beam in its waist was of 50 μm. If the total time of laser fragmentation is too short then not all of Al NPs may pass through the beam waist and therefore be fragmented. The size distribution of nanoparticles was determined using a CPS DC2400 measuring disk centrifuge. The volume of the colloidal solution injected into the centrifuge is of 50 μl. The morphology of nanoparticles was characterized with Transmission Electron Microscope (TEM) JEM-100C.

Laser fragmentation of Al NPs is accompanied by laser breakdown of the liquid and plasma formation. The breakdown starts on the NPs inside the laser beam waist. The time of laser fragmentation was found empirically to be around 20 minutes for the volume of isopropanol used in the experiments. The initial concentration of Al NPs in the liquid is another experimental parameter that was varied in the experiments in the range 100-300 μg/ml. Higher initial concentrations of Al NPs lead to scattering of the laser beam in the suspension, and the energy density of laser radiation decreases.

TEM images of both initial Al NPs and NPs after laser fragmentation in isopropanol are presented in Fig. 1. Initial Al NPs are rather large and sometimes are intercalated between each other. Laser fragmentation leads to formation of much smaller NPs. Also one can see that there is a significant amount of diffuse component around small Al NPs. This may be tentatively attributed to glassy carbon formation and incomplete pyrolysis of isopropanol.

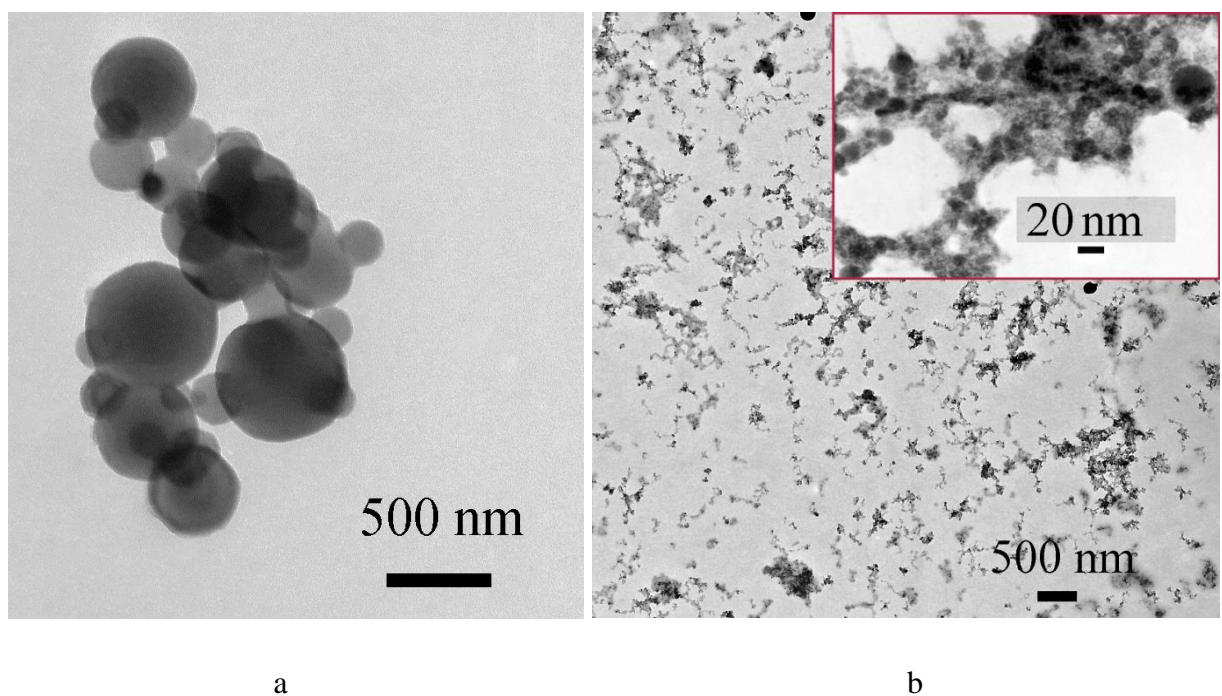

a                                                                                              b

Fig. 1. TEM images of initial Al nanoparticles (a) and nanoparticles after laser fragmentation in isopropanol (b). Scale bars denote 500 nm. The time of laser fragmentation was of 20 minutes. In the inset: enlarged view, scale bar denotes 20 nm.

Small Al NPs are embedded into this diffuse component. No oxide layer is visible within the limits of resolution of the microscope.

Distributions of the mass, number of particles, and their area against their size are presented in Fig. 2.

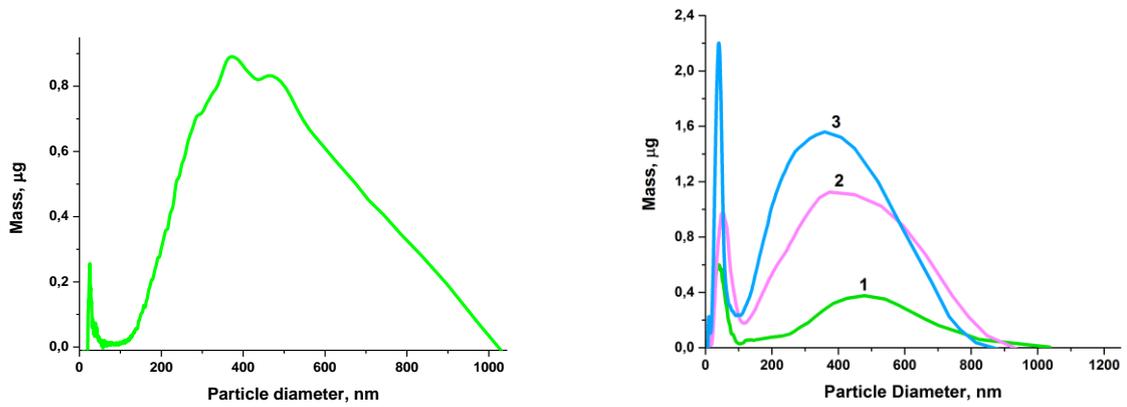

a

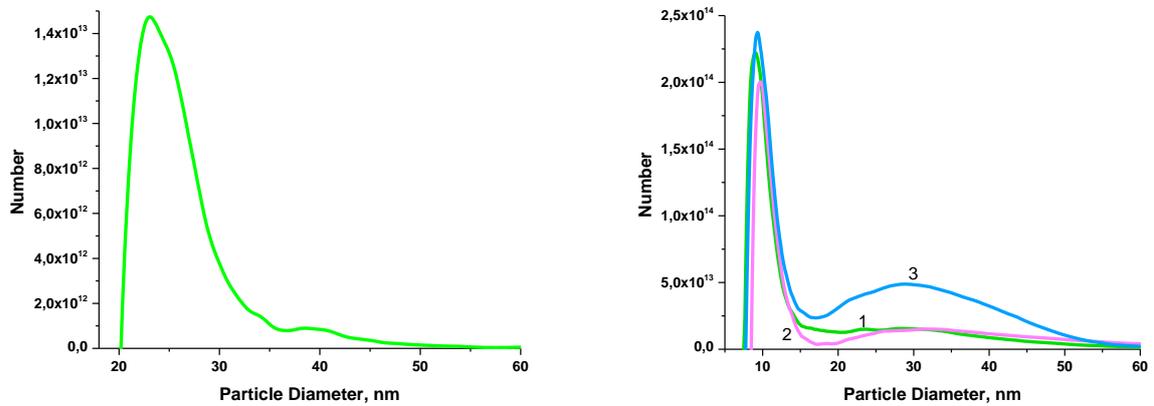

b

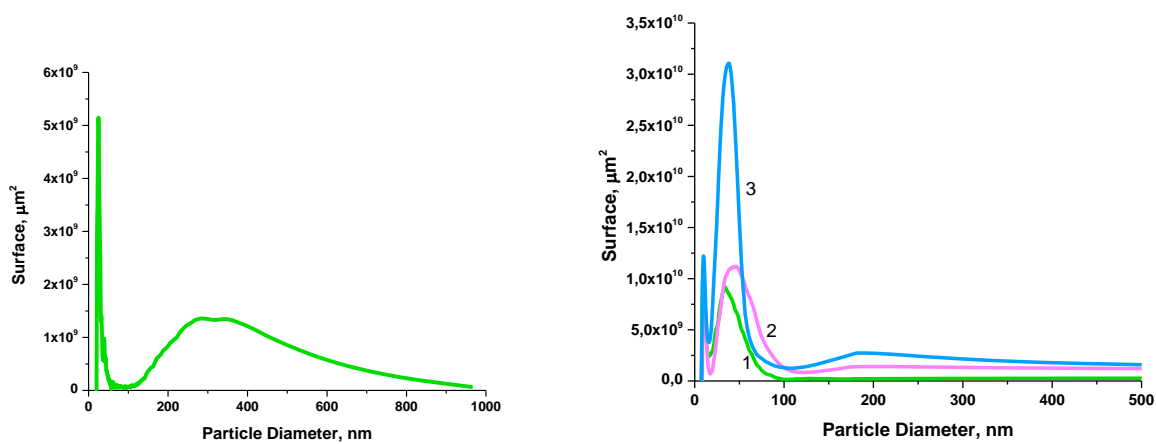

c

Fig. 2. Distribution of the mass (a), number of nanoparticles (b), and surface of nanoparticles (c) on their size at various initial concentrations of Al NPs in isopropanol. Left graph corresponds to initial Al NPs (Alex) at concentration of 100 μg/ml, right graph shows the fragmented Al NPs. Concentration of initial Al NPs is of 100 (1), 200 (2), and 300 μg/ml.

One can see that the distribution of nanoparticles on size is bimodal. This result is consistent with previous model [9]. According to it, the laser fragmentation of NPs of order of 100 nm in size proceeds through detachment of a small fragment from larger nanoparticle. Also, at the given time of laser fragmentation the fraction of relatively large NPs with size about 300 nm is still presented in the colloidal solution of higher concentration (Fig. 2, a). This means that large but already fragmented NPs should pass through the beam waist several times to be fragmented to small NPs. The majority of NPs have the size around 10 nm. NPs with the size of 20 nm give the main contribution to the specific surface.

The series of extinction spectra of Al NPs as the function of time of laser fragmentation is shown in Fig. 3.

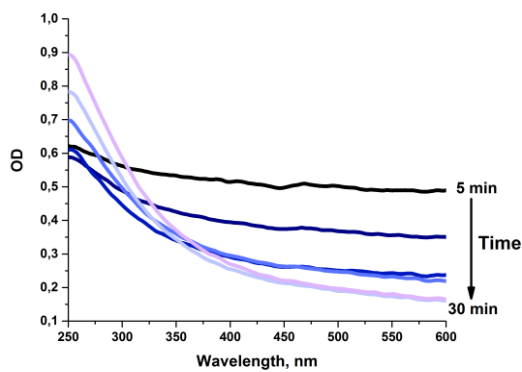

Fig. 3. Evolution of the extinction spectrum of the the colloid of Al NPs in isopropanol with time of laser fragmentation. Concentration of Alex in isopropanol of 200 μg/ml. The spectra are taken with pure isopropanol as a reference.

In the beginning of laser fragmentation the solution looks grey. The solution becomes more transparent under further fragmentation. Final solution after 30 min of fragmentation looks yellowish due to some absorption in the blue region of spectrum. Final extinction spectrum coincides with theoretical spectrum of extinction of 10 nm Al NPs in the medium with the refractive index of $H_2O$ [11]. Therefore, one can assign this spectrum to plasmon resonance of metallic Al NPs.

Thus, the efficient laser fragmentation of relatively large Al NPs into smaller NPs has been demonstrated. The surface of Al NPs increases after fragmentation by one order of magnitude. These small NPs are imbedded into diffuse halo made presumably of glassy carbon. Fragmented Al NPs can be used for synthesis of composite hydrocarbon fuels with definite content of Al nanoparticles of known size and mass. The glassy carbon formed during laser fragmentation in isopropanol is not a problem for hydrocarbon fuel, since it also burns increasing thus the heat release and temperature rise in the flame.

Acknowledgements

The research was supported by RSF (project No. 20-19-00419). This work was partly performed within the framework of National Research Nuclear University MEPhI (Moscow Engineering Physics Institute) Academic Excellence Project (Contract No. 02.a 03.21.0005). O.V. Uvarov is thanked for TEM characterization of the samples.

References

1. D.E. C. Allen, G. Mittal, C.-J. Sung, E. Toulson, T. Lee, Proc. Combust. Inst. (2011) 33, 3367–3374.
2. Jones, M., Li, C.H., Afjeh, A. et al. Experimental study of combustion characteristics of nanoscale metal and metal oxide additives in biofuel (ethanol). Nanoscale Res Lett (2011) 6, 246.

3. Valery V. Smirnov, Sergey A. Kostritsa, Vitaly D. Kobtsev, Nataliya S. Titova, Aleksander M. Starik, Experimental study of combustion of composite fuel comprising n-decane and aluminum nanoparticles, Combustion and Flame (2015), 162(10), 3554-3561.

4. Dokhan A, Price EW, Seitzman JM, Sigman RK, The effects of bimodal aluminum with ultrafine aluminum on the burning rates of solid propellants. Proc. Combus. Inst. (2002) 29, 2939–2946.

5. Jayaraman K, Anand KV, Chakravarthy SR, Sarathi R: Effect of nano-aluminium in plateau-burning and catalyzed composite solid propellant combustion. Combust Flame (2009) 156, 1662–1673.

6. E. Stratakis, M. Barberoglou, C. Fotakis, G. Viau, C. Garcia, and G. A. Shafeev, Generation of Al nanoparticles via ablation of bulk Al in liquids with short laser pulses, Optics Express (2009), 17(15), 12650.

7. G. Viau, V. Colliére, L.-M. Lacroix, G.A. Shafeev, Internal structure of Al hollow nanoparticles generated by laser ablation in liquid ethanol, Chemical Physics Letters (2011) 501, 419–422.

8. P.G. Kuzmin, G.A. Shafeev, G. Viau, M. Barberoglou, E. Stratakis, C. Fotakis, Porous nanoparticles of Al and Ti generated by laser ablation in liquids, Applied Surface Science, 258 (2012) 9283– 9287.

9. P.G. Kuzmin, G.A. Shafeev, A.A. Serkov, N.A. Kirichenko, M.E. Shcherbina, Laser-assisted fragmentation of Al particles suspended in liquid, Applied Surface Science (2014) 294, 15-19.

10. I.V. Baymler, E.V. Barmina, A.V. Simakin, G.A. Shafeev, Generation of hydrogen under laser irradiation of organic liquids, Quantum Electronics (2018) 48 (8) 738 – 742.

11. Creighton J.A., Eadon D.G., J. Chem. Soc. Faraday Trans., (1991) 87, 3881.